\documentclass[english,conference,10pt]{IEEEtran}
\usepackage[T1]{fontenc}
\usepackage[utf8]{inputenc}
\usepackage{babel}
\usepackage{amsmath}
\usepackage{amsfonts}
\usepackage{amssymb}
\usepackage{units}
\usepackage{graphicx}
\usepackage{subcaption}
\usepackage{url}
\usepackage{hyperref}
\usepackage[capitalise]{cleveref}
\crefname{equation}{}{}
\begin{document}

\title{Daala: Building A Next-Generation Video Codec From Unconventional
Technology}

\author{\IEEEauthorblockN{Jean-Marc Valin, Timothy B. Terriberry, Nathan E. Egge, Thomas Daede,\\
Yushin Cho, Christopher Montgomery, Michael Bebenita}\vspace{1.18mm}
\IEEEauthorblockA{Mozilla, Mountain View, USA}
\IEEEauthorblockA{Xiph.Org Foundation, USA}
jmvalin@jmvalin.ca}
\maketitle
\begin{abstract}
Daala is a new royalty-free video codec that attempts to compete with
 state-of-the-art royalty-bearing codecs.
To do so, it must achieve good compression while avoiding all of their
 patented techniques.
We use technology that is as different as possible from traditional
 approaches to achieve this.
This paper describes the technology behind Daala and discusses where it fits in
 the newly created AV1 codec from the Alliance for Open Media.
We show that Daala is approaching the performance level of more mature,
 state-of-the art video codecs and can contribute to improving AV1.
\end{abstract}

\section{Introduction}

Daala~\cite{DaalaWebsite} is a video codec designed to explore a set of
 atypical techniques, outlined in \cref{sec:techniques}, to avoid the patent
 thickets built around most current codecs.
Some of these techniques are new to Daala, while others already existed, but
are not used in popular standards. Although Daala is not yet
a competitive codec on its own, some of the techniques it uses are
currently being integrated in the Alliance for Open Media 
(AOM)~\cite{AOMWebsite} codec, AV1, which we discuss in~\cref{sec:AOM}.
\cref{sec:Results} presents results obtained with both Daala and AV1.

\section{Daala Techniques}
\label{sec:techniques}

Most of the techniques Daala uses have been fundamental to the design since the
initial stages of the project. Some of the techniques described below, like
lapped transforms, Overlapped Block Motion Compensation (OBMC), and non-binary
arithmetic coding, have forced the entire codec to move in a very
different direction.

Vector variables are
denoted in bold ($\mathbf{x}$) and their individual components are denoted
with indices ($x_i$). Quantized variables are denoted with a hat ($\hat{x}$).
Unless otherwise noted, $\|\mathbf{x}\|$ denotes the $L2$-norm of a vector
$\mathbf{x}$.

\subsection{Lapped Transforms}
\label{sec:lapping}

\begin{figure}
\centering
\includegraphics[width=0.8\columnwidth]{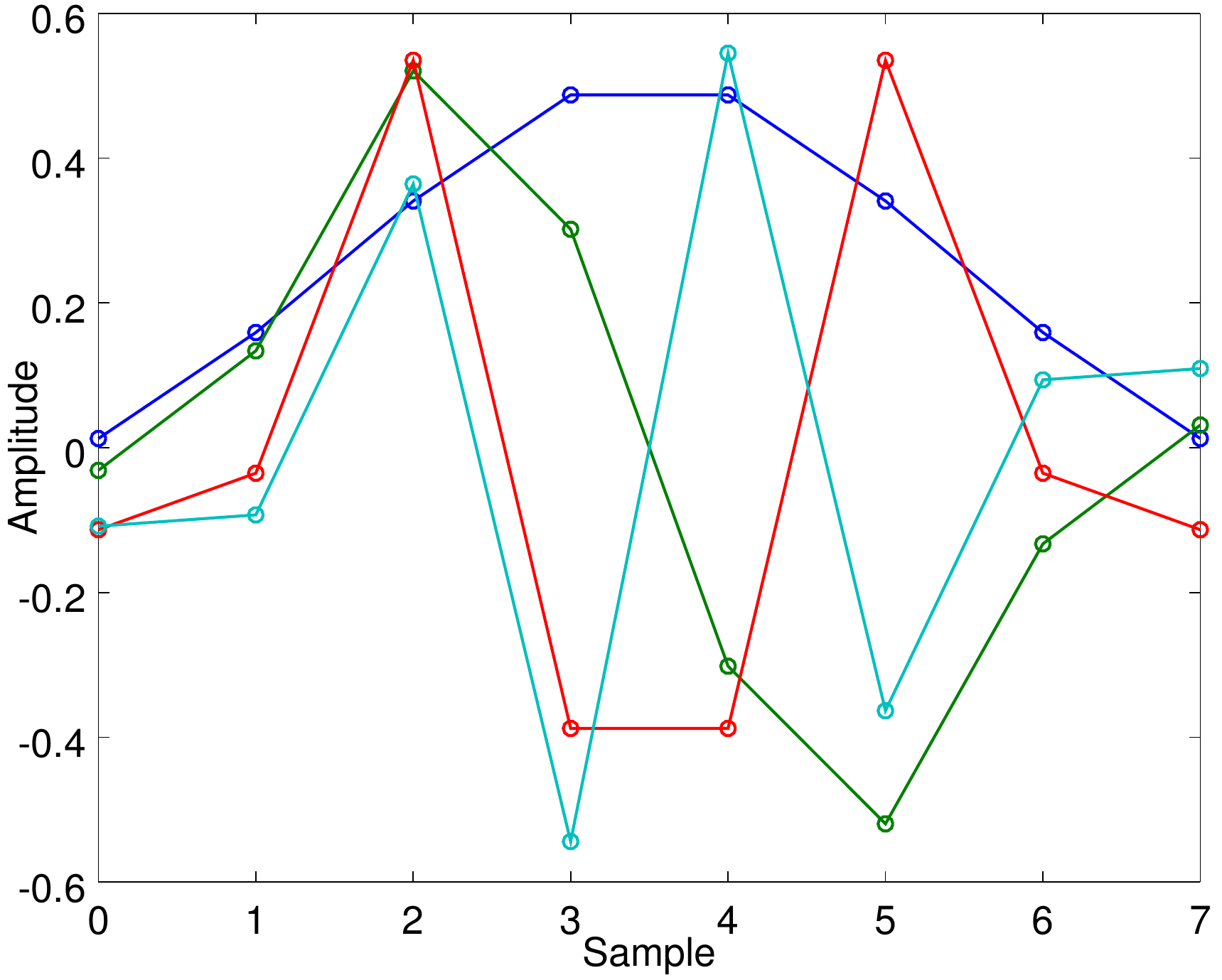}
\caption{1D synthesis basis functions for the lapped $4 \times 4$ DCT.\label{fig:basis4}}
\end{figure}

Rather than using a deblocking filter to attenuate blocking artifacts caused
by quantizing DCT coefficients, Daala uses biorthogonal lapped
transforms~\cite{MalvarS89,Tran2003}. The transform applies
a \textit{decorrelating} pre-filter to the input image before computing
the DCT and applies a \textit{deblocking} post-filter to the reconstruction
after the inverse DCT\@. Since the pre-filter is the inverse of the post-filter,
there is no need for complex adaptation of the filter strength to avoid blurring
details. As shown in \cref{fig:basis4}, the post-filter causes the basis functions
to decay smoothly at transform block edges, avoiding blocking artifacts.

Transform block sizes in Daala range from $4 \times 4$
to $64 \times 64$, based on recursive quad-tree subdivision. A
\textit{superblock} refers to the largest area of a frame on which Daala
can operate. Their size is $64\times 64$.

Although applying lapping that spans the entire width of the transform is
sometimes desirable (giving an $N \times N$ transform a support of $2N \times 2N$),
it makes block size decision using rate-distortion optimization~(RDO) intractable.
Any choice of block size affects the coding efficiency of neighboring blocks. For
this reason, Daala uses 4-point lapping for all transform sizes.

One disadvantage of lapped transforms is that it complicates intra prediction.
Because of the overlap, the pixels adjacent to the block being predicted are not
available for use in intra prediction. They cannot be reconstructed
without quantized transform coefficients from the block being predicted.
Instead of pixel-domain intra prediction, Daala uses a simple
frequency-domain intra predictor. We predict AC coefficients along horizontal
and vertical directions by directly copying a row or column of AC
coefficients from the block above and the block to the left~\cite{EggePCS}.
Some previous attempts at a more general
frequency-domain intra predictor~\cite{fdintra-demo} were ultimately abandoned,
as they failed to achieve good results with tractable complexity on large block
sizes and mixed block sizes.

\subsection{Haar DC}
\label{sec:HaarDC}

\begin{figure}
\centering
\begin{subfigure}[t]{0.49\columnwidth}
\includegraphics[width=\columnwidth]{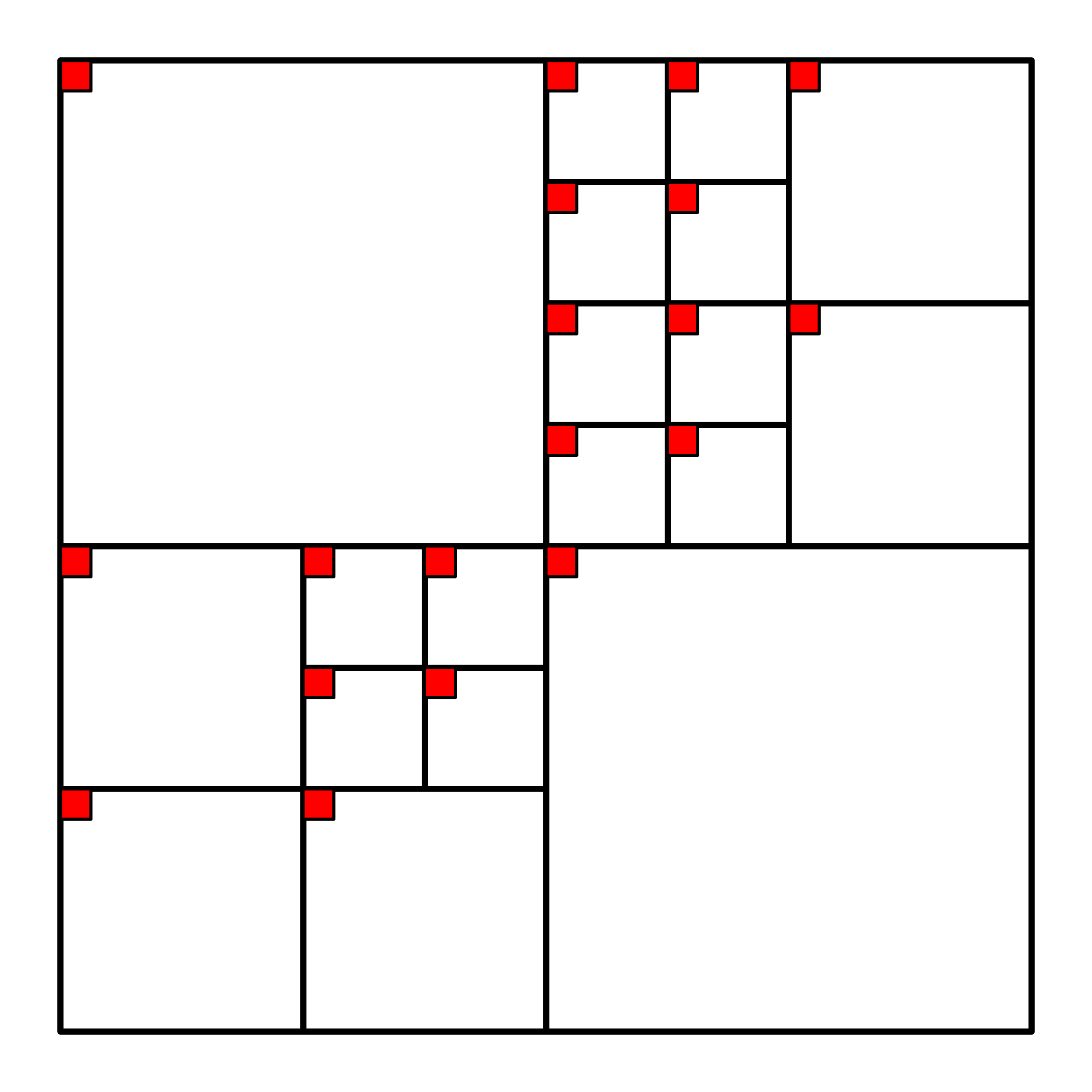}
\caption{Original DC coefficients before Haar DC}
\end{subfigure}
\begin{subfigure}[t]{0.49\columnwidth}
\includegraphics[width=\columnwidth]{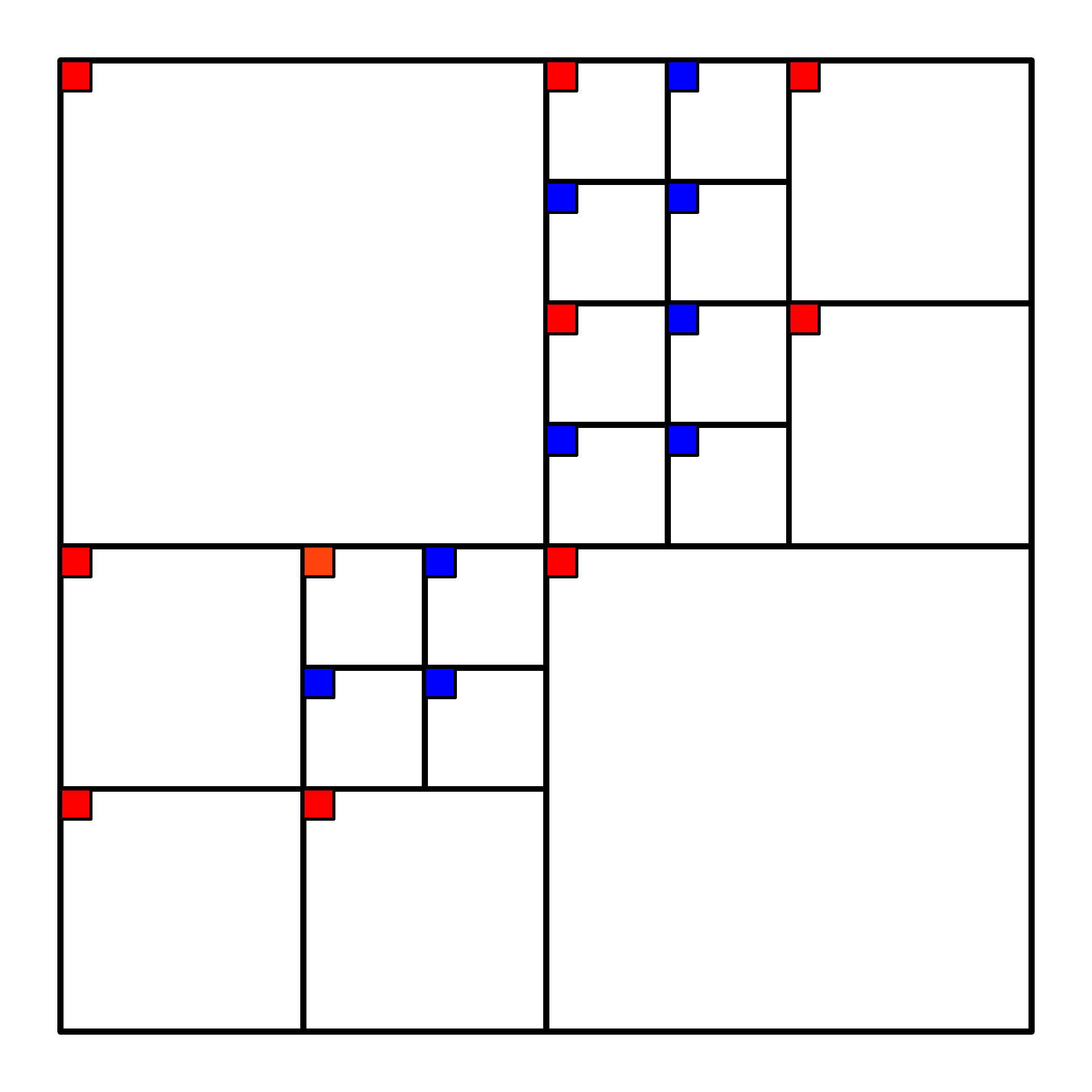}
\caption{DC coefficients from $8\times 8$ blocks are combined}
\end{subfigure}
\begin{subfigure}[t]{0.49\columnwidth}
\includegraphics[width=\columnwidth]{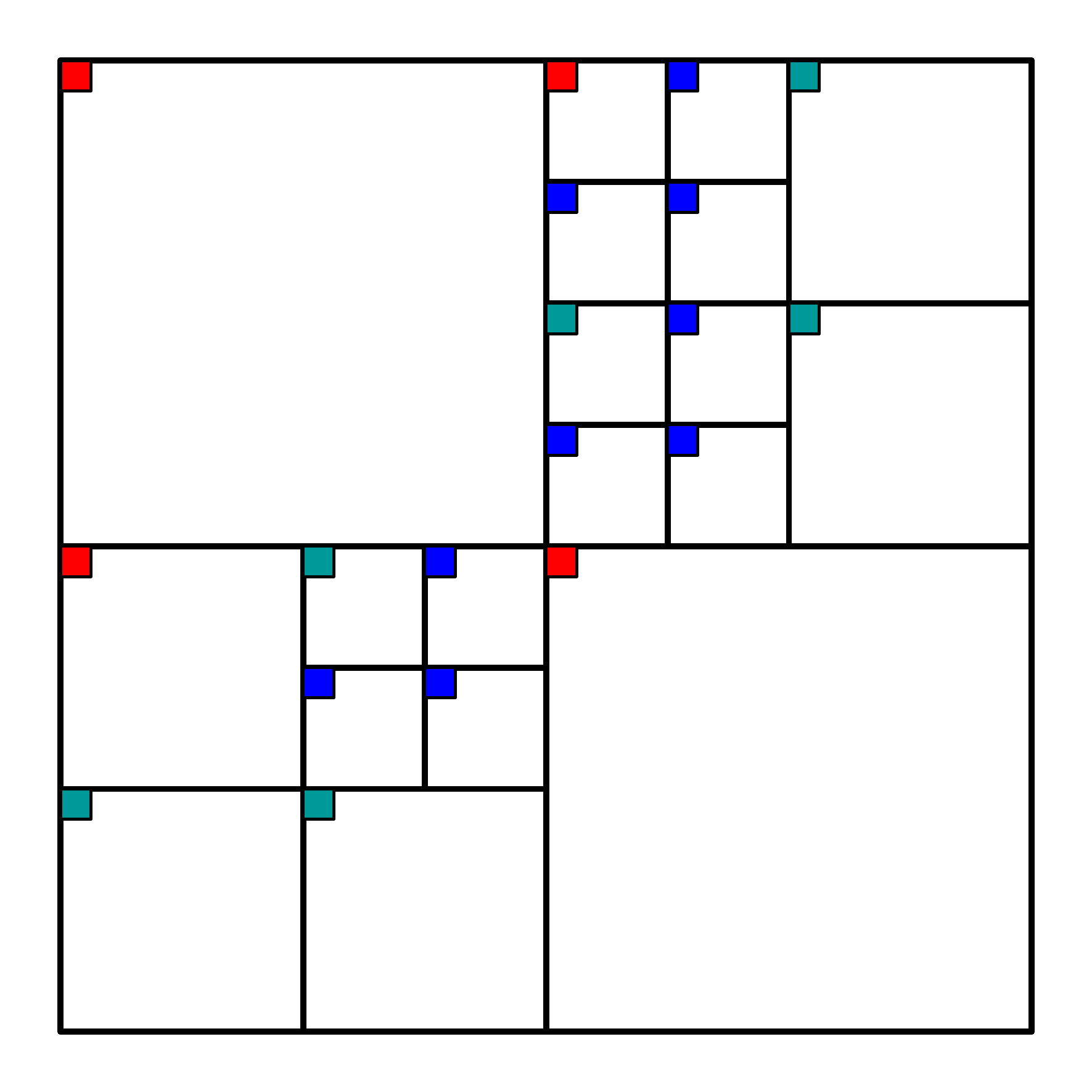}
\caption{DC coefficients from $16\times 16$ sub-blocks are combined}
\end{subfigure}
\begin{subfigure}[t]{0.49\columnwidth}
\includegraphics[width=\columnwidth]{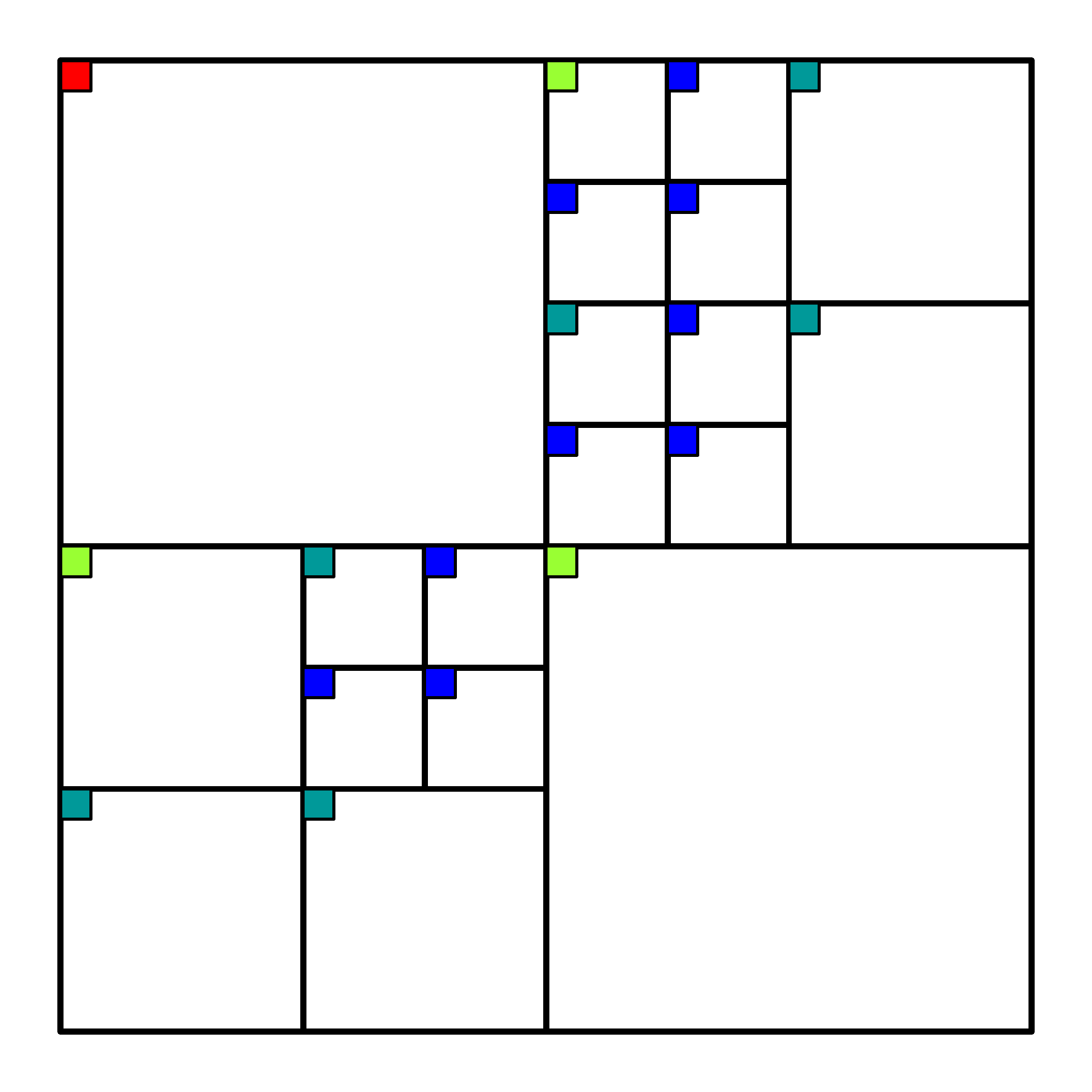}
\caption{DC coefficients from $32\times 32$ sub-blocks are combined}
\end{subfigure}
\caption{Example of applying Haar DC over three levels on a $64\times 64$
superblock subdivided into transform blocks of $8 \times 8$, $16 \times 16$,
and $32 \times 32$. At each step (a) to (d), the remaining DC coefficients
are shown in red\label{fig:haardc}}

\end{figure}

Instead of using intra prediction for DC coefficients in keyframes, they
are further transformed with a 2D Haar wavelet.
Since Daala transform blocks are always split as quad-trees, the transform is
applied bottom-up, recursively, up to the level of the corresponding superblock,
as shown in \cref{fig:haardc}.
At each level, four DCs are combined into four Haar coefficients: one horizontal,
one vertical, one diagonal, and one new DC corresponding to a larger block size.

The highest level ($64\times 64$) DC is predicted as a linear combination of the
neighboring superblock DC coefficients: left, top-left, top, and top-right.
The prediction coefficients are trained on an image database and constrained to
sum to unity. The
(non-DC) horizontal and vertical Haar coefficients are predicted from
co-located horizontal and vertical coefficients at a larger scale. This slightly
reduces bitrate in smooth areas.

\subsection{Multi-Symbol Entropy Coder}
\label{sec:EntropyCoder}

Most recent video codecs encode information using binary arithmetic
coding, meaning that each symbol can only take two values. The Daala
range coder supports up to 16 values per symbol, making it possible
to encode fewer symbols~\cite{derfTools}. This is equivalent to
coding up to four binary values in parallel and reduces serial dependencies,
allowing hardware implementations to use lower clock rates, and thus less
power.

The range coder itself is based on a multiply-free approximation presented
in~\cite{stuiver1998piecewise}, called piecewise integer mapping. The original
approximation overestimates probabilities for symbols at the end of the alphabet, and underestimates
probabilities at the beginning of the alphabet. Daala does this the
other way around, since $0$ is frequently the most probable symbol.
Overestimating its probability leads to lower approximation overhead than
underestimating it without reordering the alphabet.

Piecewise integer mapping replaces the multiplication and division of a traditional
arithmetic coder with a subtraction, a minimum operation, and an addition. This
allows it to work with any input probability distribution, without requiring
that its total be normalized to a power of two. That lets Daala use traditional
frequency counts to model probabilities, instead of having to extend the
table-based schemes usually used by binary coders to larger alphabet sizes.
Probabilities can typically be updated with just one or two SIMD instructions
in software, and are similarly cheap to update in parallel in hardware. The
overall cost is a bitrate overhead of around $1\%$ in practice,
which is comparable to CABAC~\cite{MSW03}.

We are currently exploring approaches for reducing this overhead without
sacrificing throughput or modeling efficiency with hardware implementers in the
AOM\@.

\subsection{Overlapped Block Motion Compensation}
\label{sec:OBMC}

Because hard block edges are expensive to code with lapped transforms, we
want to avoid creating blocking artifacts in the motion-compensated
prediction. For this we use overlapped block motion compensation
(OBMC)~\cite{OBMC}. We use an adaptive grid of Motion Compensation
(MC) blocks ranging from $8\times 8$ up to $64\times 64$, in order to scale to
high-resolution content. This grid is completely independent of the transform
blocks, and the block sizes of one do not impose any constraints on the block
sizes of the other.
We also experimented with $4\times 4$ MC blocks, but they were rarely used in
practice, and actually caused a small quality regression at equal bitrate with
the encoder used at the time.

The use of variable-sized MC blocks requires a blending scheme that maintains
continuity between neighboring regions of different sizes. One approach, used
by codecs like Dirac~\cite{Dirac}, is to fix the overlap size at the largest
overlap allowed by the smallest motion-compensated block (8~pixels, in this
case). This makes a large
block equivalent to a group of smaller blocks with the same motion vector.
However, this can create low-passing artifacts in a predictable grid pattern.
These can be visually annoying, and require many bits to correct because of
their locality. Unlike with transform blocks, there is no pre-filter in the
encoder to compensate for the blending done in the decoder.
Zhang \textit{et al.} showed that splitting large blocks,
but only as far as necessary to prevent their blending windows from
overlapping more than one neighboring block, improved quality~\cite{ZAS98}.
However, they implemented this as a post-process to a non-overlapped
block-based motion search, and did not incorporate RDO\@.

Instead, Daala structures its grid as a 4-8 mesh~\cite{DWSMAM97} to ensure the
sizes of neighboring MC blocks differ by no more than a factor of two. This
makes it easy to design OBMC blending windows that both span an entire block
and ensure continuity across block size changes. Although R-D optimal block
size decisions with this data structure are still NP-hard, there is a fast
dynamic programming approximation that achieves good results in
practice~\cite{Bal01}. The details of the 4-8 grid structure, the blending
windows, and the dynamic programming algorithm are outlined in~\cite{OBMC}.

\subsection{Perceptual Vector Quantization}
\label{sec:PVQ}

Most video codecs subtract a motion-compensated reference frame
from the input frame to compute a residual, then transform and code it
using scalar quantization. Instead, Daala uses gain-shape vector quantization.
This encodes the signal as a vector by splitting it into a magnitude (gain),
and direction (shape). Most importantly, in order to ensure the gain represents
the amount of energy in the original signal, the motion-compensated
reference is never subtracted from the input frame. Instead, Daala uses that
reference to build a transformation that makes the input easier to code.
This technique is called Perceptual
Vector Quantization (PVQ)~\cite{valin2015spie}.

Perceptual Vector Quantization originates from the pyramid vector quantizer
previously used for music in the Opus audio codec~\cite{ValinAES}.
The pyramid vector quantizer is also a gain-shape quantizer that Opus uses to
ensure that the signal energy is always conserved. Using a gain-shape
quantizer in a video codec is more complicated, since it must also take
into account a predictor. While we could quantize the scalar difference
between the prediction and the input, conserving the energy of that
difference is perceptually meaningless.

Rather than attempting to encode the difference, we derive a Householder
reflection from the predictor that makes the input easier to encode. Let
$\mathbf{x}$ be the input and $\mathbf{r}$ be the prediction.
We construct a Householder reflection plane
\begin{equation}
\mathbf{v} = \frac{\mathbf{r}}{\|\mathbf{r}\|} + s\mathbf{e}_m\ ,
\end{equation}
where $\mathbf{e}_m$ is a unit vector along dimension $m$ and $s = \pm1$.
The values of $m$ and $s$ are arbitrary, but to maximize
numerical stability, we typically choose $m$ to be the position of the
largest absolute value in $\mathbf{r}$ and $s$ to be the sign of that value.

We then apply the reflection to the input vector
$\mathbf{x}$ to produce the reflected vector $\mathbf{z}$:
\begin{equation}
\mathbf{z} = \mathbf{x} - 2\frac{\mathbf{x}^T\mathbf{v}}
{\mathbf{v}^T\mathbf{v}}\mathbf{v}\ .
\end{equation}
When the input is similar to the prediction itself, the direction of the
reflected vector $\mathbf{z}$ is close to the axis $-\mathbf{e}_m$. To take
advantage of that fact, we express it as
\begin{equation}
\mathbf{z} = g\left(-s\cos\theta + \mathbf{u}\sin\theta\right)\ ,
\end{equation}
where $g$ is the magnitude of $\mathbf{z}$ (and thus also the magnitude of
$\mathbf{x}$), $\mathbf{u}$ is a unit vector with no component along the
$\mathbf{e}_m$ direction, and $\theta$ is the angle between $\mathbf{r}$ and
$\mathbf{x}$ (a meaningful parameter that represents the similarity between
the prediction and the input). Since the Householder reflection is orthonormal,
it follows that
\begin{equation}
\theta = \arccos\frac{\mathbf{x}^T\mathbf{r}}
                   {\left\|\mathbf{x}\right\|\left\|\mathbf{r}\right\|}\ .
\end{equation}

We code the unit vector $\mathbf{u}$ using a spherical quantizer derived
from the pyramid vector quantizer~\cite{Fischer1986}:
\begin{equation}
\mathbf{u}=\frac{\mathbf{y}}{\left\|\mathbf{y}\right\|}\ ,
\end{equation}
with
\begin{equation}
\mathbf{y} \in \mathbb{Z}^N : \left\|\mathbf{y}\right\|_{L1} = K \land y_m=0\ ,
\end{equation}
where the number of \textit{pulses} $K$ controls the size of the codebook.

The encoder quantizes $g$ and $\theta$ and encodes them in the bitstream along
with the integer vector $\mathbf{y}$ (excluding $y_m$ which is $0$). The codebook
size $K$ is determined only from $g$ and $\theta$ and does not need to be
transmitted. Since the decoder has access to the prediction vector $\mathbf{r}$,
it can compute the reflection vector $\mathbf{v}$ without the need to transmit
$m$ and $s$. There are $N-2$ degrees of freedom to code in
$\mathbf{y}$, and two more for $g$ and $\theta$. Thus we still code parameters
with a total of $N$ degrees of freedom. The main difference from scalar
quantization is that two of the
coded values have a perceptual meaning: $g$ is the amount of contrast and
$\theta$ is the amount of deviation from the prediction. By coding $g$
as a parameter, it is easier to preserve the amount of contrast than by
coding only DCT coefficients.

In practice, the vectors $\mathbf{x}$ and $\mathbf{r}$ are transform
coefficients rather than pixel values. This requires an extra forward DCT
in both the encoder and the decoder since the input and the prediction need
to be transformed separately. Only the AC coefficients are coded using PVQ,
and for transform blocks larger than $4\times 4$, the AC coefficients are divided into multiple
\textit{bands}, where each band is coded separately. This allows us
to control the contrast separately in each octave and orientation.

PVQ also allows us to take into account masking effects with no
extra signaling. Since the gain is explicitly signaled, we can make
the quantization resolution depend on the gain:
\begin{equation}
E\left\lbrace \left\| \mathbf{x} - \hat{\mathbf{x}} \right\|^2 \right\rbrace
\propto g^{2\alpha}\ , 0 \leq \alpha \leq 1\ ,
\end{equation}
where $\alpha=0$ behaves like a standard linear scalar quantizer and
$\alpha=1$ produces a constant relative error like in the Opus audio codec.
Daala uses $\alpha = \nicefrac{1}{3}$. To achieve this, we quantize the
companded gain
\begin{equation}
\gamma = g^{1-\alpha}\ ,
\end{equation}
giving finer resolution to smaller gains and coarser resolution to larger
gains.

The decoder always decodes the quantized companded gain $\hat{\gamma}$
first. From there it can compute the quantization step size for $\theta$ as
\begin{equation}
Q_\theta = \frac{\beta}{\hat{\gamma}}\ ,
\end{equation}
where $\beta = \frac{1}{1-\alpha}$.

We determined the size of the codebook $K$ through curve
fitting~\cite{valin2015spie} to be
\begin{equation}
K = \frac{\hat{\gamma}\sin\hat{\theta}}{\beta}\sqrt{\frac{N+2}{2}}\ .
\label{eq:K_nonrobust}
\end{equation}
The formulation in~\cref{eq:K_nonrobust} is not robust to packet loss when
the gain is predicted. If there are errors in the prediction, the decoder may
obtain the wrong $K$ and decode the wrong number of symbols. Fortunately, by making the $\sin{\hat{\theta}}
\simeq \hat{\theta}$ approximation and substituting $\hat{\theta} =
Q_\theta\hat{\tau}$, where $\hat{\tau}$ is the quantization index of the angle, we obtain
\begin{equation}
K = \hat{\tau} \sqrt{\frac{N+2}{2}}\ .
\end{equation}

\subsection{Chroma from Luma (CfL) Prediction}
\label{sec:CFL}
Although the use of $\mathrm{Y'C_BC_R}$ reduces the correlation across planes
compared to RGB, the chroma planes $\mathrm{C_B}$ and $\mathrm{C_R}$ and the
luma plane $\mathrm{Y'}$ are still often locally correlated. Edges in
chroma tend to
align very well with edges in luma, with only the amount of contrast (gain) differing.
PVQ's separation of signals into a gain and a shape makes it especially easy
to predict chroma planes from the luma plane. Daala's chroma from luma (CfL)~\cite{egge2015spie}
prediction uses the luma transform coefficients as the prediction vector
$\mathbf{r}$ directly. The only complication is that we need to code a sign for the prediction,
since the luma plane and chroma plane coefficients may be negatively correlated.
We also do not predict the gain of chroma from the gain of luma.

\subsection{Directional Deringing Filter}
\label{sec:deringing}

Like other transform codecs, Daala can cause ringing artifacts around edges.
We use a directional deringing filter to attempt to eliminate the ringing without
blurring the image. Unlike HEVC's Sample-Adaptive Offset (SAO)~\cite{HEVC-SAO},
the Daala deringing filter is not based on classifying pixels and applying per-class
offsets. Instead, it is an outlier-robust directional filter that smooths the
neighborhood of pixels while preserving edges.

Let $x\left(n\right)$ denote a 1-dimensional signal and $w_{k}$
denote filter tap weights. We define a linear finite impulse response (FIR)
filter with unit DC response as
\begin{equation}
y\left(n\right)=\frac{1}{\sum_{k}w_{k}}\sum_{k}w_{k}x\left(n+k\right)\ ,\label{eq:FIR1}
\end{equation}
which can alternatively be written as
\begin{equation}
y\left(n\right)=x\left(n\right)+\frac{1}{\sum_{k}w_{k}}\sum_{k,k\neq0}w_{k}\left[x\left(n+k\right)-x\left(n\right)\right]\ .\label{eq:FIR2}
\end{equation}
The main advantage of expressing a filter in the form of~\cref{eq:FIR2}
is that the normalization term $\frac{1}{\sum_{k}w_{k}}$ can be approximated
relatively coarsely without affecting the unit gain for DC\@. This makes
it easy to use small integers for the weights $w_{k}$.

The disadvantage of linear filters for removing ringing artifacts
is that they tend to also cause blurring. To reduce the amount of
blurring, Daala uses a ``conditional replacement filter'' to exclude
the signal taps $x\left(n+k\right)$ that would cause blurring and
replace them with $x\left(n\right)$ instead. It determines this
by whether $x\left(n+k\right)$ differs from $x\left(n\right)$ by
more than a threshold $T$. This makes the FIR filter in~\cref{eq:FIR2}
into a conditional replacement filter:
\begin{equation}
y\left(n\right)=x\left(n\right)+\frac{1}{\sum_{k}w_{k}}\sum_{k,k\neq0}w_{k}R\left(x\left(n+k\right)-x\left(n\right),T\right)\ ,\label{eq:CRF}
\end{equation}
where
\begin{equation}
R\left(x,T\right)=\left\{ \begin{array}{ll}
x & ,\ \left|x\right|<T\\
0 & ,\ \mathrm{otherwise}
\end{array}\right.\ .
\end{equation}

To further reduce the risk of blurring the decoded image, we apply the
conditional replacement filter along the main direction of the edges
in each $8\times 8$ block of the reconstructed image. We determine the
direction from the decoded image (no side information is transmitted) by
analyzing each $8\times 8$ block as described
in~\cite{ValinDeringing}. For each $8\times 8$ block, the decoder determines which
of eight different directions best represents the content of the block.
The search can be efficiently implemented in SIMD\@. We
apply a 7-tap conditional replacement filter along the detected direction to
each pixel in the $8\times 8$ block.

To further reduce ringing in very smooth regions of the image, we apply a
second filter to combine multiple output values of the
first filter. The second filter is applied either vertically or horizontally
-- in the direction most orthogonal to the one used in the first filter.
For a 45-degree direction, we apply the second filter horizontally to reduce
hardware line buffer requirements. The combined effect of the two filters is a
separable deringing filter that covers a total of 35~pixel taps.

\begin{figure}
\centering
\begin{subfigure}[t]{0.5\columnwidth}
\includegraphics[width=\columnwidth]{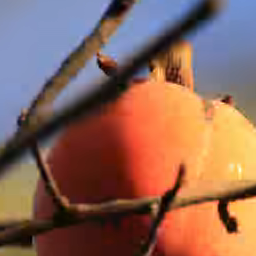}
\includegraphics[width=\columnwidth]{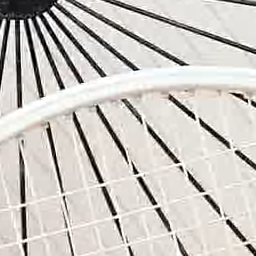}
\caption{Without deringing filter}
\end{subfigure}\begin{subfigure}[t]{0.5\columnwidth}
\includegraphics[width=\columnwidth]{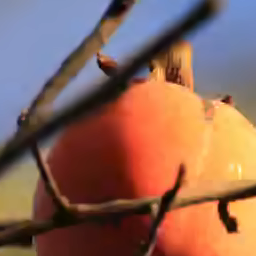}
\includegraphics[width=\columnwidth]{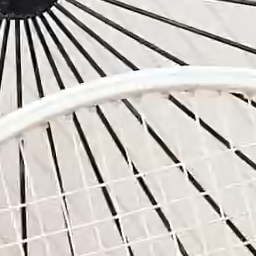}
\caption{With deringing filter}
\end{subfigure}
\caption{Effect of the deringing filter on two images with sharp edges.\label{fig:deringing}}
\end{figure}

The deringing filter only requires a minimal amount of signaling. We send a
global threshold for the entire a frame, and signal one of six adjustment
factors (including an \textit{off} setting) for each superblock. With entropy coding, the
cost of the signaling generally averages 2~bits per superblock, or 128~bytes for
a 1080p keyframe. For predicted frames, we don't apply the deringing filter to
$8\times 8$ blocks where no coefficients are coded, and no adjustment factor
is coded for superblocks where no $8\times 8$ block is filtered, further
reducing the amount of signaling.
\cref{fig:deringing} shows the effect of the deringing filter at low bitrates.

\section{Alliance for Open Media AV1 codec}
\label{sec:AOM}

The recently-formed Alliance for Open Media (AOM) is currently specifying
AV1, a royalty-free video codec. The initial development is based on
technology from three existing codecs: Google's VP9~\cite{VP9Spec} codec,
Cisco's Thor~\cite{ThorDraft} codec, and the Daala codec presented
here. For this reason, some of the techniques used in Daala are currently
being considered for inclusion in AV1.

The deringing filter described in \cref{sec:deringing} is already
fully integrated in AV1 and has been shown to reduce bitrate by around 2\%
at equal quality. AV1 also includes Thor's constrained lowpass
filter~(CLPF)~\cite{CLPFDraft} that attenuates ringing, but with a more
limited effect and a lower complexity than Daala's deringing filter.

We are also evaluating the multi-symbol entropy coder (\cref{sec:EntropyCoder})
for use in AV1. Probability distributions in AV1 are currently fixed or
explicitly signaled. In this case, we can avoid the piecewise integer mapping
overhead by using probabilities whose denominator is a power of two. We expect
that using adaptive distributions will produce larger gains than the overhead
introduced by the mapping approximation, but using this effectively requires
revisiting how every symbol is coded.

PVQ (\cref{sec:PVQ}) is in the early stages of experimentation within AV1 and
is by far the most invasive of the Daala techniques under consideration.
Integration requires many changes to the bitstream, as well as the addition
of a forward transform on the
prediction itself. No results are available yet, but should PVQ be
included in AV1, we would also be able to add CfL (\cref{sec:CFL}).

Lapped transforms (\cref{sec:lapping}) and OBMC (\cref{sec:OBMC}) are 
\textit{not} being considered for inclusion
in AV1. Both these techniques have far-reaching interactions with the
other coding techniques and would essentially require a complete redesign
of the codec. Considering that Haar~DC (\cref{sec:HaarDC}) is mostly
needed to compensate for the
lack of pixel-domain prediction, it also is not currently being considered for AV1.

\section{Results and Discussion}
\label{sec:Results}

\begin{figure*}
\centering
\begin{subfigure}[t]{0.44\textwidth}
\includegraphics[width=\columnwidth]{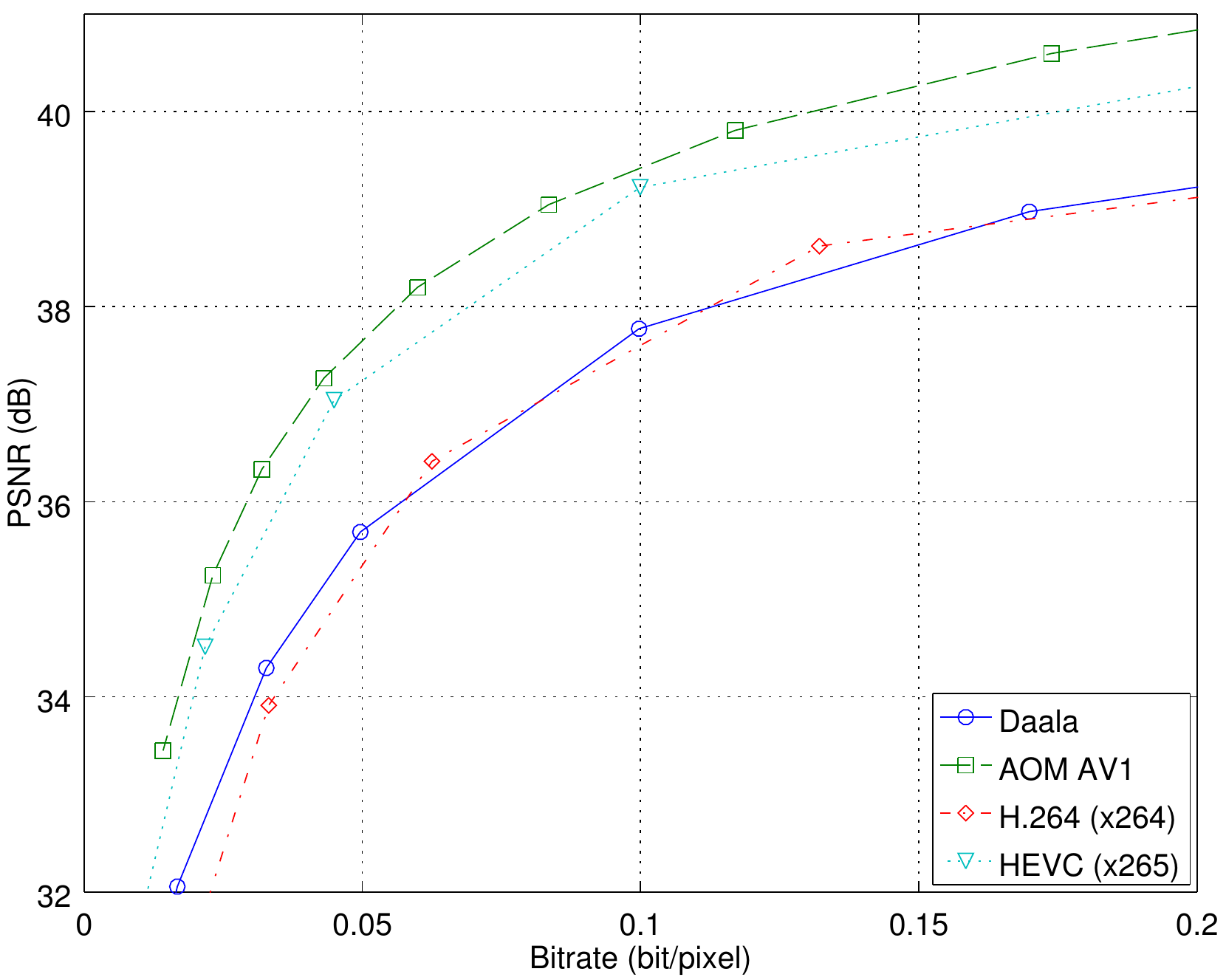}
\caption{PSNR}
\end{subfigure}\hspace{0.12\columnwidth}\begin{subfigure}[t]{0.44\textwidth}
\includegraphics[width=\columnwidth]{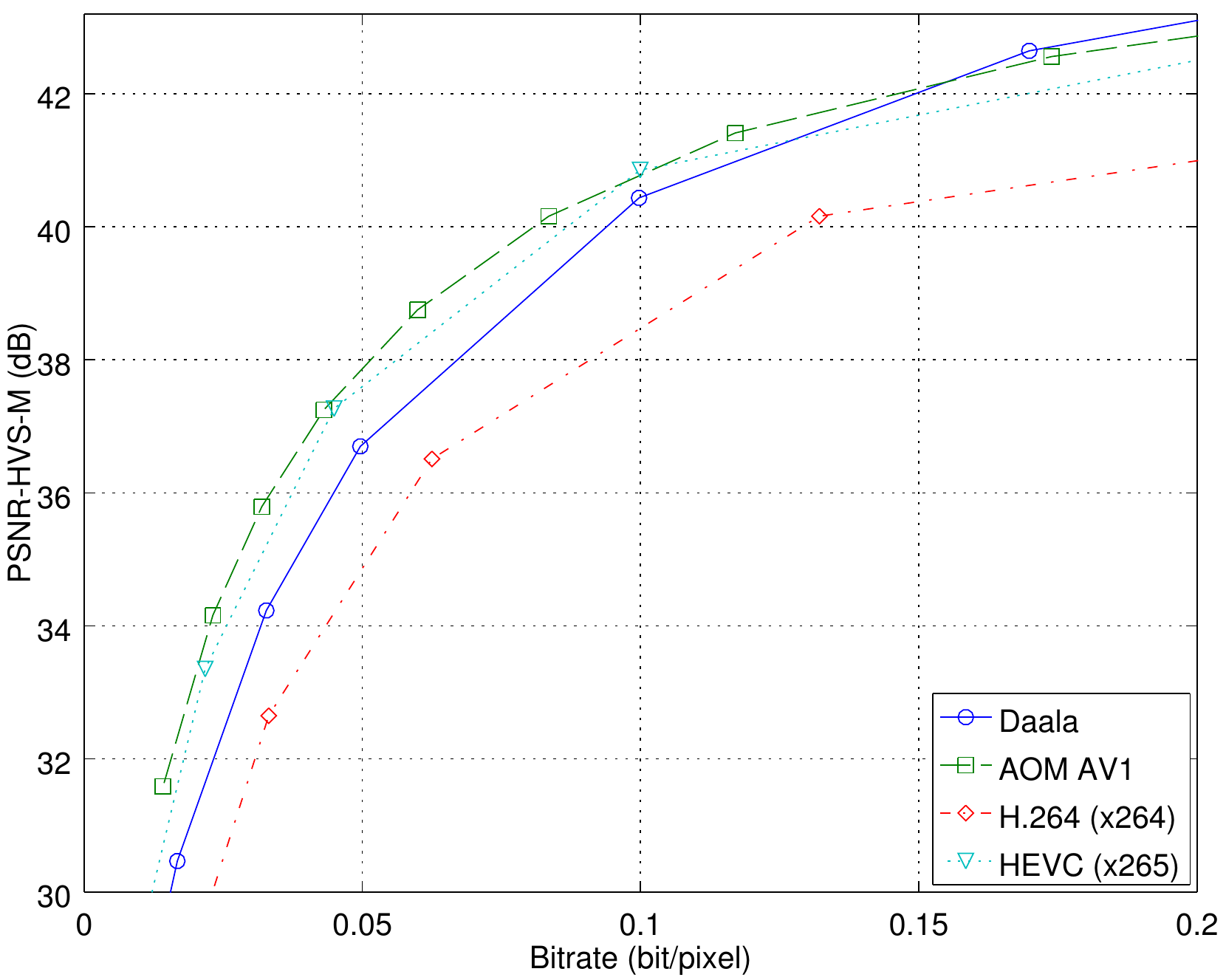}
\caption{PSNR-HVS-M}\label{fig:psnrhvsm}
\end{subfigure}
\begin{subfigure}[t]{0.44\textwidth}
\includegraphics[width=\columnwidth]{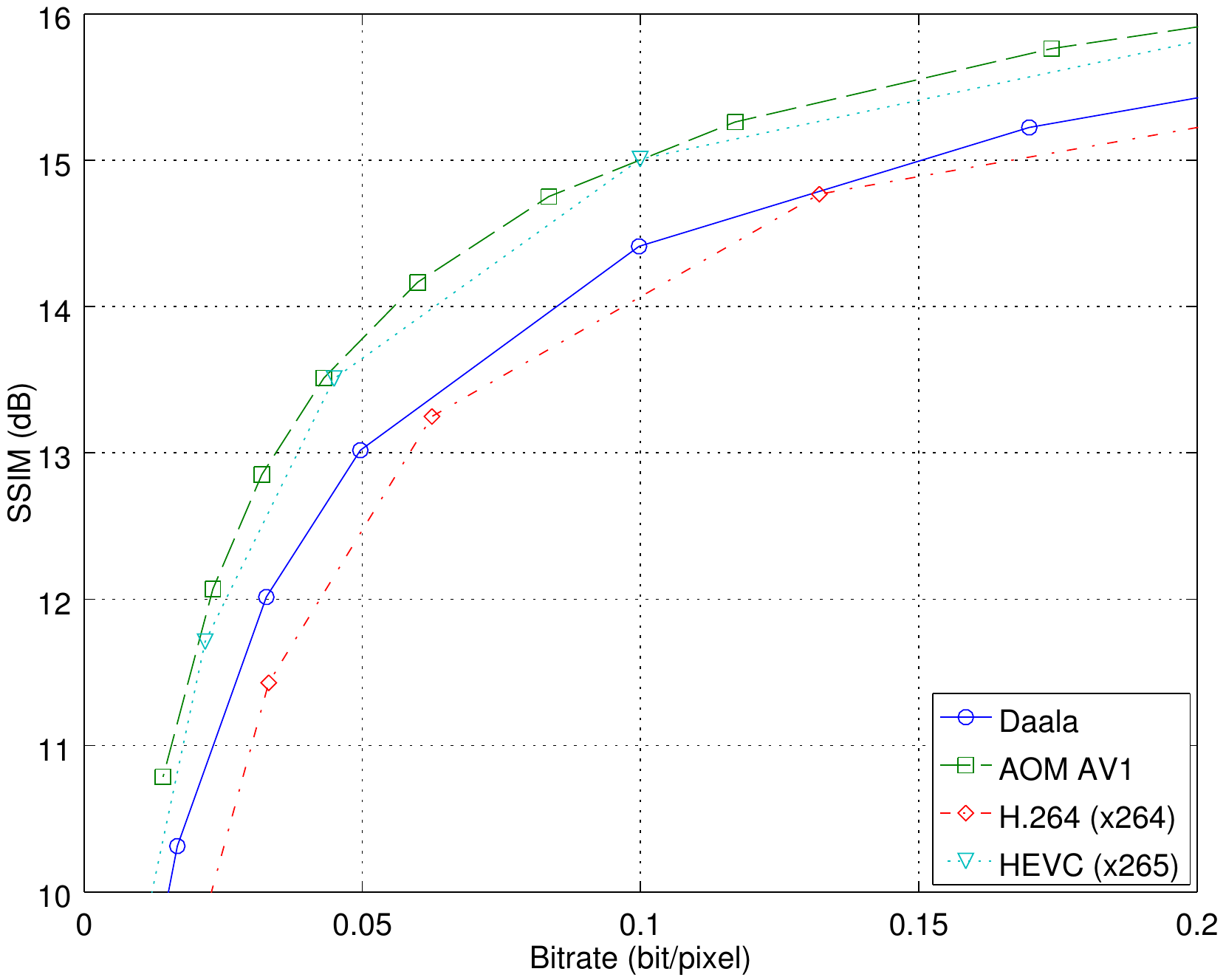}
\caption{SSIM}
\end{subfigure}\hspace{0.12\columnwidth}\begin{subfigure}[t]{0.44\textwidth}
\includegraphics[width=\columnwidth]{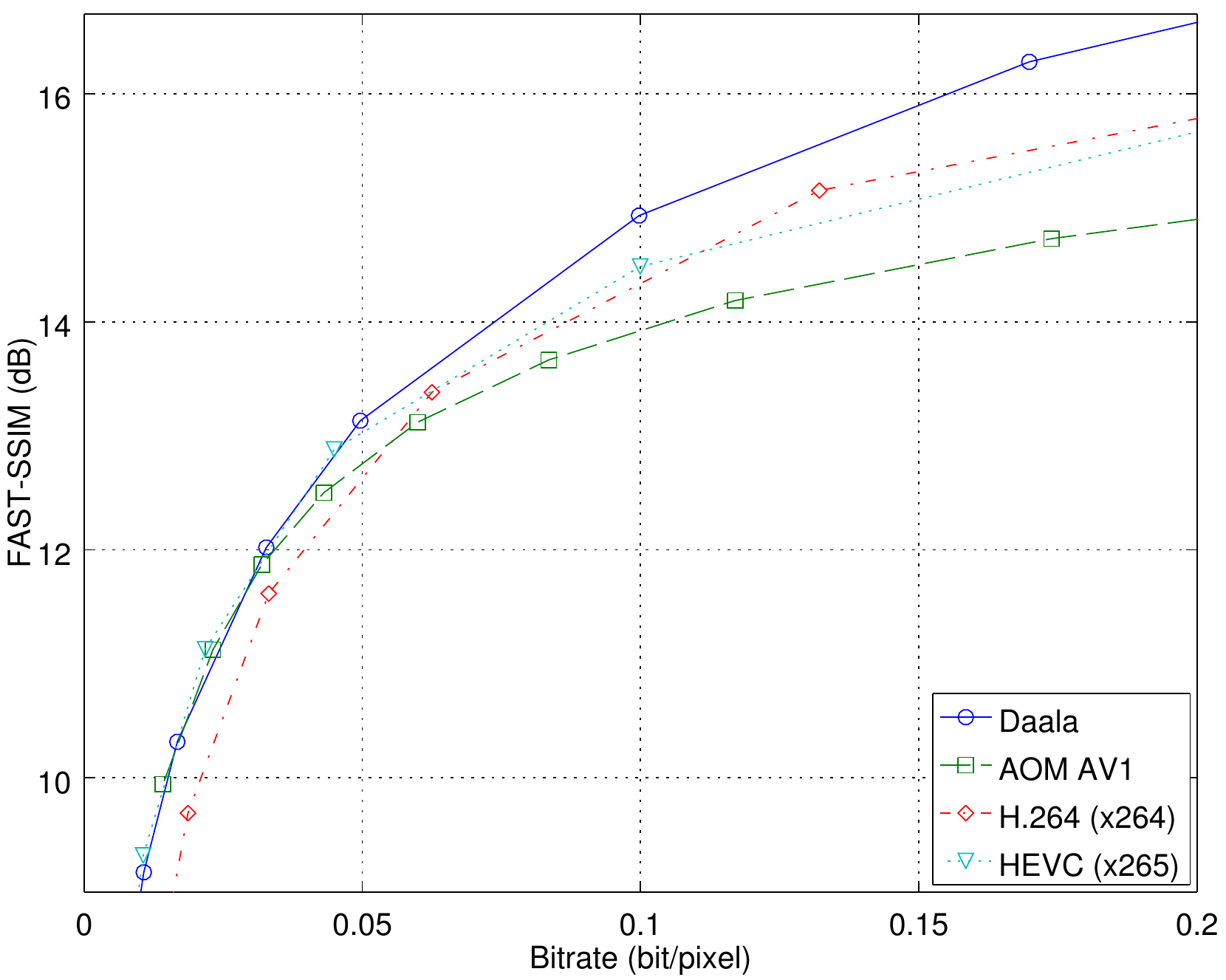}
\caption{FAST-SSIM}
\end{subfigure}
\caption{Comparison between Daala, HEVC and AV1 based on objective metrics.}\label{fig:metrics}
\end{figure*}

We have tested Daala on the \texttt{ntt-short-1} test set
using the Are We Compressed Yet?~\cite{AWCY} testing infrastructure.
We use four different objective metrics for the comparison: PSNR,
PSHR-HVS-M~\cite{PSHRHVSM}, SSIM, and FastSSIM~\cite{FastSSIM} (a low-complexity
version of multiscale SSIM). We compare
Daala\footnote{Git version d55fff01 from April 25th, 2016 with command-line options: -k 1000 -v}
with the AV1 encoder\footnote{Git version 337b23a5 from April 7th, 2016 with deringing enabled and command-line options: --ivf --frame-parallel=0 --tile-columns=0 --auto-alt-ref=2 --cpu-used=0 --passes=2 --threads=1 --kf-min-dist=1000 --kf-max-dist=1000 --lag-in-frames=25 --end-usage=q --cq-level=}, the x264 H.264
encoder\footnote{Command-line options: --preset placebo --min-keyint 1000 --keyint 1000 --no-scenecut --crf=},
and the x265 HEVC encoder\footnote{Version 1.6 with command-line options: --preset slow --frame-threads 1 --min-keyint 1000 --keyint 1000 --no-scenecut --crf=}.

The results in \cref{fig:metrics} show that Daala is generally better than
H.264, and slightly worse than HEVC and AV1. Based on this authors' informal
evaluation, the subjective performance of Daala is close to what the PSNR-HVS-M
results show in \cref{fig:psnrhvsm}. Qualitatively, the Daala artifacts tend to differ
greatly from most other video codecs. Daala tends to perform more poorly on sharp
details and edges, while retaining more texture in low contrast regions. This is in part due
to PVQ activity masking, but remains true even without activity masking.

Considering that most of the technology used in Daala is either new or unproven
in the context of video codecs, we consider these results to be
encouraging. For example, Daala's results presented here do not use B-frames,
while all of the other codecs do use B-frames or Alt-Refs to significantly
improve coding efficiency.
Moreover, these results have been steadily improving over the past
two years, suggesting that Daala may be a viable approach to royalty-free codecs
in the longer term. Some of the techniques presented in this paper will make their
way into the AV1 codec, while others will require more time to mature before
being used in a video coding standard.

\bibliographystyle{IEEEtran}
\bibliography{daala}

\end{document}